\documentclass[english,aps,pra, twocolumn, floatfix, showpacs, superscriptaddress]{revtex4-1}
\usepackage{amsmath}
\usepackage{amsfonts}
\usepackage{graphicx}
\usepackage{color}
\usepackage[colorlinks=true,urlcolor=blue,citecolor=blue,linkcolor=blue]{hyperref}
\usepackage[sort&compress]{natbib}
\usepackage{epstopdf}

\newcommand{\figref}[1]{Fig.~\ref{#1}}

\begin{document}

\title{Photonic currents in driven and dissipative resonator lattices}

\begin{abstract}
Arrays of coupled photonic cavities driven by external lasers represent a highly controllable setup to explore photonic transport. 
In this paper we address (quasi)--steady states of this system that exhibit photonic currents introduced by engineering driving and dissipation. We investigate two approaches: in the first one, photonic currents arise as a consequence of a phase difference of applied lasers and in the second one, photons are injected locally and currents develop as they redistribute over the lattice. Effects of interactions are taken into account within a mean--field framework. In the first approach, we find that the current exhibits a resonant behavior with respect to the driving frequency. Weak interactions shift the resonant frequency toward higher values, while in the strongly interacting regime in our mean--field treatment the effect stems from multiphotonic resonances of a single driven cavity. For the second approach, we show that the overall lattice current can be controlled by incorporating few cavities with stronger dissipation rates into the system. These cavities serve as sinks for photonic currents and their effect is maximal at the onset of quantum Zeno dynamics.
\end{abstract}

\author{Thomas Mertz}
\affiliation{Institut f\"ur Theoretische Physik, Goethe-Universit\"at,
60438 Frankfurt/Main, Germany}
\author{Ivana Vasi\'c}
\affiliation{Scientific Computing Laboratory, Institute of Physics Belgarde, University of Belgrade, 11080 Belgrade, Serbia}
\affiliation{Institut f\"ur Theoretische Physik, Goethe-Universit\"at,
60438 Frankfurt/Main, Germany}
\author{Michael J.~Hartmann}
\affiliation{Institute of Photonics and Quantum Sciences, Heriot-Watt University, Edinburgh, EH14 4AS, United Kingdom}
\author{Walter Hofstetter}
\affiliation{Institut f\"ur Theoretische Physik, Goethe-Universit\"at,
60438 Frankfurt/Main, Germany}
\maketitle

\section{Introduction}

Understanding the transport properties of photons in different media is a prerequisite for future applications, for example in quantum  information processing. This subject has been addressed from various perspectives \cite{Carusotto}. As one notable example we mention successful experimental realizations of photonic topological insulators, where emerging edge states provide robust transport channels \cite{MITphotons, Rechtsman, Hafezi, Carusotto, LeHur}. Forthcoming experiments with arrays of coupled photonic cavities \cite{Hartmann2, Houck, Carusotto} are expected to feature strong interactions on a single--photon level. The latest theoretical and experimental progress in this direction is summarized in two recent review papers \cite{Noh, Hartmannreview}. Transport measurements will be the most natural first experiments to carry out in these systems in order to explore how interactions affect the propagation of photons. First experimental results in this direction are already available \cite{Abbarchi, Raftery}. 

Theoretically,  arrays of coupled photonic cavities can be described by the Bose--Hubbard model \cite{Hartmann2, Houck, Carusotto}. However, photonic cavities exhibit dissipation due to intrinsic loss rates, which has to be compensated by driving the system with an external laser. Instead of equilibrium properties, stationary states that arise from the interplay of driving and dissipation are thus more naturally studied in this open quantum system  \cite{Carusotto2, Tomadin, Hartmann, Nissen, Jin, Boite, Boite, Boite2, Pizorn, Laptyeva, Biella, Naether}. The aim of our study is to explore steady states of the dissipative--driven two--dimensional Bose--Hubbard model which exhibit finite photonic currents, and are generated by  engineering the driving and dissipation. In particular, we will analyze how the emerging photonic currents are affected by the externally controllable parameters, such as intensity and frequency of the external laser pump, the loss rates and the physical parameters of the underlying Bose--Hubbard model.

We note that transport measurements in cold atomic systems  \cite{Bloch} have been reported recently \cite{Brantut, Krinner, Labouvie, Labouvie2} and that some of our conclusions may apply to corresponding bosonic systems of cold atoms as well. Different possibilities to control stationary flows of cold atoms by dissipation have been theoretically addressed in Refs.~\cite{Brazhnyi, Zezyulin, Kreibich, Single, Haag}.

The structure of the paper is the following. The model we consider is described in Sec.~II, where we also introduce two setups, which lead to stationary states with finite currents. In Sec.~III we briefly outline the theoretical methods we employ in this work. In Sec. IV we explore properties of the currents first in the non--interacting limit, then at weak interactions and finally in the regime of strong interactions, where we use the Gutzwiller mean--field approximation.
In the end we summarize our main conclusions and outline open questions.

\section{The model}

\begin{figure}
	\includegraphics[width=\linewidth]{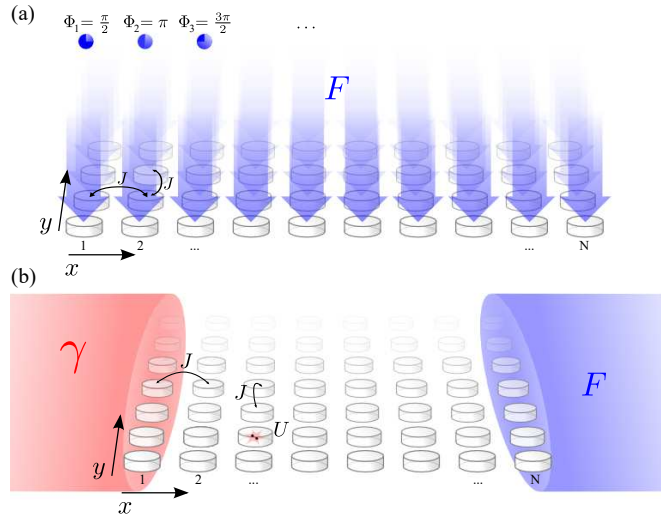}
	\caption{(a) Sketch of the phase imprinting setup (\ref{eq:phaseimprint}). (b) Sketch of the source--drain setup (\ref{eq:sourcedrain}). Throughout this paper we assume translational invariance in $y$-direction, where all sites along the $y$-axis behave in the same way.}
	\label{fig:sketch}
\end{figure}

 We study transparency in the dissipative--driven  photonic Bose-Hubbard model, which describes the dynamics of photonic/polaritonic excitations in coupled cavity arrays, see Fig.~\ref{fig:sketch} for a sketch of our setup. The key parameters of the Bose-Hubbard model  are the hopping amplitude $J$ and the on-site interaction $U$. The driving of the system via local excitation by external lasers can be described by $F_l^* a_l \exp\left(i \omega_L t\right) + \mathrm{h.c.}$, where the amplitudes $F_l$ are set by the laser intensity and $a_l$ are the bosonic annihilation operators on site $l$.  We describe the system in the co-rotating frame, by applying the  unitary transformation $U(t)=\exp(i\omega_L t \sum_l n_l), n_l=a_l^{\dagger} a_l$. This transformation leads to an additional chemical potential-like term proportional to the detuning $\Delta = \omega_L - \omega_C$ of the laser frequency with respect to the cavity mode $\omega_C$. The effective Hamiltonian of the model is \cite{Boite2, Biella}
\begin{eqnarray}
 \mathcal{H} &=& -\Delta \sum_l a_l^{\dagger} a_l-J\sum_{<l,j>}\left(a_l^{\dagger} a_j+a_j^{\dagger} a_l\right)\nonumber\\
 &+&\frac{U}{2}\sum_l n_l (n_l - 1)+\sum_l \left( F_l a_l^{\dagger} + F_l^* a_l \right),
 \label{eq:ham}
\end{eqnarray}
where the sum over $<l,j>$ indicates that we only take into account tunneling between nearest--neighbor sites of the square lattice.
In addition to the Hamiltonian time evolution we consider one-body loss described by a Lindblad master equation. The equations of motion for the density operator $\rho$ of the dissipative model are given by
\begin{equation}
 i \frac{d \rho}{d t} = \left[\mathcal{H}, \rho\right]+\mathcal{L} \rho,
 \label{eq:eom}
\end{equation}
where we set $\hbar =1$. The dissipator $\mathcal{L}$ is
\begin{equation}
\mathcal{L} \rho =  i \sum_l \frac{\gamma_l}{2}\left(2 a_l \rho a_l^{\dagger}-a_l^{\dagger} a_l \rho-\rho a_l^{\dagger} a_l\right),
\label{eq:diss1}
\end{equation}
where $\gamma_l$ is the local dissipation rate.

In order to quantify the transparency of the material we calculate the (local) current density $j$, which is derived from the lattice continuity equation and provides a measure for the photon transport through the system.
The current $j_{lj}$ between sites $l$ and $j$ is given by
\begin{equation}
 j_{lj} = -i J\left(  a_j^{\dagger} a_l -  a_l^{\dagger} a_j\right),
 \label{eq:current}
\end{equation}
and is the main quantity commonly used to describe transport in other lattice systems, as for example in \cite{Benenti, Prosen2, Mendoza}. 
From the experimental side, the two--point correlations $\langle  a_j^{\dagger} a_l \rangle$ have already been measured in superconducting circuits \cite{Filipp}, implying that photonic bond currents may be directly accessible in forthcoming experiments. Another possibility for probing properties of a photonic flow is through a local loss of photons, which will be explained in the next section.

In our study we will investigate the photonic transport in the regime of finite local bosonic coherences given by $|\langle a_l \rangle |$.
In this case it is reasonable to approximate the expectation value of the bond current from Eq.~(\ref{eq:current}) by
\begin{equation}
 \langle j_{lj} \rangle \approx -i J\left( \langle a_j^{\dagger} \rangle \langle a_l \rangle - \langle a_l^{\dagger} \rangle \langle a_j\rangle\right),
\end{equation}
where the local expectation values $\langle a_l \rangle$ are calculated  within a mean--field approximation. From the last equation it follows that the current is directly related to the phase ordering of the complex expectation values $\langle a_l \rangle$ of lattice nearest neighbors, and that it is enhanced by strong bosonic coherences $|\langle a_l \rangle| $. In the following we consider different spatial distributions of the dissipation rates $\gamma_l$ and the driving amplitudes $F_l$ in order to find an optimal regime where the steady states exhibit maximal bond currents. Due to the symmetry of the considered setups we assume translational invariance in $y$-direction, where all sites along the $y$-axis behave in the same way. In this case there is no  current in $y$ direction and the indices $l$ and $j$ label $x$ coordinates of the lattice sites, see Fig.~\ref{fig:sketch}. 

One possibility to realize steady states exhibiting a finite bond current is by engineering suitable phases of the coherent driving terms $F_l$
\begin{equation}
 F_l^{\mathrm{PI}} = F \exp\left(i \Phi_l\right), \quad \Phi_l = \Phi^{\mathrm{PI}} l, \quad \gamma_l^{\mathrm{PI}} = \gamma_b,
 \label{eq:phaseimprint}
\end{equation}
that will be imprinted onto phases of $\langle a_l \rangle$, thus providing the finite current.
This setup has been introduced in Ref.~\cite{Hartmann} and throughout the paper we designate it as phase imprinting (PI), Fig.~\ref{fig:sketch}(a).
A second experimentally relevant protocol that leads to steady states with currents uses drives that inject photons into the lattice locally e.g.~by shining laser light on one side only, Fig.~\ref{fig:sketch}(b). Steady states in the presence of homogeneous dissipation in a one--dimensional lattice have been explored recently in such systems \cite{Biella}. In particular, stronger loss rates at the opposite lattice side should serve as photonic sinks 
 \begin{equation}
        F_l^{\mathrm{SD}} = F \delta_{l, N}, \quad \gamma_l^{\mathrm{SD}} = \gamma \delta_{l, 1} + \gamma_b,
        \label{eq:sourcedrain}
\end{equation}
thus providing for a stable photonic flow. In both Eqs.~(\ref{eq:phaseimprint}) and (\ref{eq:sourcedrain}), the index $l$ stands for the site position along the $x$-axis and there is no explicit dependence on the site position in $y$-direction.
The aim of our study is to explicate how the emerging current intensity $j$ is set by the laser amplitude $F$ and intrinsic loss rates $\gamma_l$, as well as by the parameters of the underlying Hamiltonian in Eq.~(\ref{eq:ham}).

We note that the onset of particle currents in a bosonic system naturally raises questions about superfluidity in a dissipative--driven system \cite{Carusotto3, Janot, Ruiz, Larre}. A definite answer can be provided by studying how the presence of defects modifies the photonic flow or by analyzing asymptotics of long-range correlations in the system. These questions will be addressed in future work.

\section{Methods}

In the non--interacting limit $U=0$ we solve the exact equations of motion for the expectation values $\phi_l = \langle a_l \rangle = \mathrm{Tr} \rho a_l$:
\begin{equation}
 i \frac{d \phi_l}{d(tJ)} = -\frac{\Delta}{J} \phi_l-\sum_{\langle l,j\rangle} \phi_j+\frac{F_l}{J}-i\frac{\gamma_l}{2 J} \phi_l,
\end{equation}
where $\langle l, j \rangle$ denotes summation over nearest--neighbor sites of the site $l$.
We consider a two--dimensional lattice with $N$ sites in $x$-direction and  translational invariance in $y$-direction implemented using periodic boundary conditions, and the index $l$ labels $x$ coordinates of the lattice sites. In this notation we have for example $\sum_{\langle l,j\rangle} \phi_j = 2 \phi_l+\phi_{l-1}+\phi_{l+1} $.
 
Using vector notation
$\vec{ \phi} = \left(\phi_{1},\ldots, \phi_{N}\right)^{T} $, $\vec{ F} = (F_{1}, \ldots, F_{N})^{T} $
the steady state solution can be written as \cite{Carusotto, Ozawa}
\begin{equation}
  \phi_l = -M^{-1}_{lj} F_j/J,
 \label{eq:inverse}
\end{equation}
where $M$ is a $N\times N $ matrix with elements
\begin{equation}
 M_{lj} = \left(-2-\Delta/J-i \gamma_l/(2 J)\right) \delta_{l,j}-\delta_{l-1,j}-\delta_{l+1,j}.
 \label{eq:matrixm}
\end{equation}
To simplify the notation, spatial indices will be omitted $ j_{lj} \rightarrow j$ from now on whenever the current throughout the lattice is constant and we implicitly assume the current between two nearest neighbors in $x$-direction.

At high densities, provided for example by strong driving, and for weak $U$, the interaction term may be treated at the mean--field level leading to nonlinearities for the $\phi_l$ in their equations of motion:
\begin{equation}
 i\frac{d \phi_l}{d(tJ)} =-\frac{\Delta}{J} \phi_l -\sum_{\langle l,j\rangle} \phi_j+\frac{U}{J} |\phi_l|^2 \phi_l+\frac{F_l}{J}-i\frac{\gamma_l}{2 J} \phi_l.
 \label{eq:GP}
\end{equation}

To get an estimate of effects of quantum fluctuations on the mean--field predictions, we follow the approach described in Ref.~\cite{Hartmann}. Using a Fourier transform
$a_{l_x,l_y} = \frac{1}{\sqrt{N_x N_y}} \sum_{\vec{k}} e^{-i\left(k_x l_x+k_y l_y\right)} B_{\vec{k}}$
we rewrite the Hamiltonian (\ref{eq:ham}) as
\begin{eqnarray}
&&\mathcal{H} = \sum_{\vec{k}}\omega_{\vec{k}} B_{\vec{k}}^{\dagger} B_{\vec{k}}+\sqrt{N_x N_y} F \left(B_{\Phi^{\mathrm{PI}},0}+B_{\Phi^{\mathrm{PI}},0}^{\dagger}\right)\nonumber\\
 &+&\frac{U}{2 N_x N_y} \sum_{\vec{k}_1, \vec{k}_2, \vec{k}_3, \vec{k}_4} \hspace{-5mm} \delta_{\vec{k}_1 + \vec{k}_2 + 2\pi (z, p), \vec{k}_3 + \vec{k}_4} B_{\vec{k}_1}^{\dagger} B_{\vec{k}_2}^{\dagger} B_{\vec{k}_3} B_{\vec{k}_4},\nonumber
\end{eqnarray}
where $\omega_{\vec{k}} = -\Delta - 2 J \left(\cos k_x+\cos k_y\right)$, and $z$ and $p$ are integers.
 In the next step, we expand operators around the mean--field solution as
\begin{equation}
 B_{\vec{k}} = \sqrt{N_x N_y} \beta \delta_{k_x, \Phi^{\mathrm{PI}}} \delta_{k_y, 0} + b_{\vec{k}},
\end{equation}
where $\left|\beta\right|^2 = n^{\mathrm{PI}} $ is the mean -- field density.
By taking into account fluctuations up to the second order we obtain an effective quadratic Hamiltonian
\begin{eqnarray}
 \widetilde{\mathcal{H}}&=& \sum_{\vec{k}}\hspace{-1mm}\left[\left(\omega_{\vec{k}}+2 n^{\mathrm{PI}} U\right) b_{\vec{k}}^{\dagger} b_{\vec{k}}+\frac{U}{2}\left( \beta^{*2} b_{\vec{k}} b_{\vec{kk}}
 +\beta^2 b_{\vec{k}}^{\dagger}  b_{\vec{kk}}^{\dagger}\right) \right]\nonumber, 
\end{eqnarray}
with $kk_x = 2 \pi z + 2 \Phi^{\mathrm{PI}} - k_x, kk_y = k_y $.
From the stationarity condition $ \frac{d }{d t} \langle b_{\vec{k}}^{\dagger} b_{\vec{k}}\rangle = 0, \quad \frac{d }{d t} \langle b_{\vec{k}} b_{\vec{kk}}\rangle = 0,$
we find closed--form equations for the second order moments
\begin{eqnarray}
 &i& U \beta^{*2} \langle b_{\vec{k}} b_{\vec{kk}}\rangle -i U \beta^2 \langle b_{\vec{k}}^{\dagger} b_{\vec{kk}}^{\dagger}\rangle-\gamma_b \langle b_{\vec{k}}^{\dagger} b_{\vec{k}}\rangle = 0, 
 \\ [0.2cm]
 -&i& \left(\omega_{\vec{k}}+\omega_{\vec{kk}}+4 n^{\mathrm{PI}} U\right) \langle b_{\vec{k}} b_{\vec{kk}}\rangle-\gamma_b \langle b_{\vec{k}} b_{\vec{kk}}\rangle \nonumber\\
  &&-i U \beta^2 \left(\langle b_{\vec{k}}^{\dagger} b_{\vec{k}}\rangle+\langle b_{\vec{kk}}^{\dagger} b_{\vec{kk}}\rangle+1\right)= 0,
\end{eqnarray}
that finally yield for $ m(\vec{k}) =\langle b_{\vec{k}}^{\dagger} b_{\vec{k}}\rangle$
\begin{equation}
 m(\vec{k}) =\frac{2  \left(U n^{\mathrm{PI}}\right)^2}{\left(\left(\omega_{\vec{k}}+\omega_{\vec{kk}}\right)^2+4 n^{\mathrm{PI}} U\right)^2+\gamma_b^2-4  \left(U n^{\mathrm{PI}}\right)^2}.\nonumber
\end{equation}
Fluctuation effects are quantified by the ratio 
\begin{equation}
 m/n^{\mathrm{PI}} = \sum_{\vec{k}} m(\vec{k})/(n^{\mathrm{PI}} N_x N_y)
 \label{eq:qfqf}
\end{equation}
and the expansion up to second order in the fluctuations can be expected to be a good approximation as long as this ratio remains small, $m/n^{\text{PI}} \ll 1$.

 When addressing the limit of strong interactions, we restrict our description to the well-established bosonic Gutzwiller approximation \cite{Tomadin, Boite, Vidanovic}, where only local correlations are taken into account. The time--dependent variational Gutzwiller mixed state is a product of local mixed states. In other words the total density operator in the Gutzwiller approximation is given by a direct product of density operators $\rho_i$ on the individual sites:
\begin{equation}
	\rho_{\mathrm{GW}}(t)= \prod_{\otimes l} \rho_l(t) = \prod_{\otimes l} \sum_{m,n <N_c} c^l_{nm}(t) \vert n \rangle_l \langle m \vert_l.
	\label{eq:gw}
\end{equation}
In our calculations we truncate the dimension of the local Hilbert space for every site at a finite value $N_c = 10$, which we choose large enough so that our results are independent of the choice of the cut-off. The accuracy and limitations of this approximation in describing dissipative systems have been discussed in Ref.~\cite{Kordas}. In brief, by comparing Gutzwiller results with exact calculations on small lattices it is found that the method describes local quantities accurately, but it underestimates phase coherence between different sites. However, it is expected that the accuracy of the method improves as the lattice coordination number increases.

Projecting the Lindblad equation \eqref{eq:eom} onto the local occupation number bases we obtain equations of motion for the variational coefficients of the Gutzwiller state, which are $N\times N_c$ coupled first order differential equations:
\begin{eqnarray}
\imath \frac{d c_{nm}^l(t)}{dt}&=& \left.\eta_l \sqrt{n} c_{n-1,m}^l+\eta_l^* \sqrt{n+1} c_{n+1,m}^l\right.\nonumber\\
&-&\left.\eta_l \sqrt{m+1} c_{n,m+1}^l-\eta_l^* \sqrt{m} c_{n, m-1}^l\right.\nonumber\\
&+& i \gamma_l \sqrt{n+1} \sqrt{m+1} c_{n+1,m+1}^l\nonumber\\
&+&\left(\frac{U}{2} (n (n-1)-m(m-1))\right.\nonumber\\&-&\left.\Delta (n-m)- i \frac{\gamma_l}{2} (n+m)\right)\,c_{n, m}^l,
\label{eq:rtgw}
\end{eqnarray}
where $\eta_l = F_l- J \sum_{\langle l, j \rangle} \phi_j $ takes into account the contribution of nearest--neighbor sites and the external driving term.
After preparing the system in an initial state we propagate the equations of motion simultaneously to describe the subsequent non-equilibrium dynamics.
We chose here to investigate the steady state solutions by observing the long-time dynamics of the system.

\section{Results}

In the following we present properties of photonic currents for the setups defined in Eqs.~(\ref{eq:phaseimprint}) and (\ref{eq:sourcedrain}).

\subsection{Phase imprinting}

In the non--interacting limit of the setup shown in Fig.~\ref{fig:sketch}a, phases of the coherent driving terms $F_l$ translate into phases of $\phi_l$ according to Eq.~(\ref{eq:inverse}) as
\begin{equation}
 \phi_l = -\sum_k \frac{1}{\varepsilon_k-\Delta-i\frac{\gamma_b}{2}} k_l \sum_j k^*_j F_j, 
 \label{eq:phi_non-int}
\end{equation}
where $\varepsilon_k$ and $|k\rangle$ are eigenfrequencies and eigenmodes of 
\begin{equation}
H^0_{lj}  =-2J \delta_{l,j}-J\delta_{l-1,j}-J\delta_{l+1,j},
\label{eq:h0}
\end{equation}
and we keep in mind that we work in a co--rotating frame. For a lattice obeying periodic boundary conditions in both $x$- and $y$-direction, we find homogeneous steady states with density  
\begin{equation}
 n^{\mathrm{PI}} = \frac{F^2}{\left(2 J \left(1+\cos \Phi^{\mathrm{PI}}\right) + \Delta\right)^2+\frac{\gamma_b^2}{4}},
\end{equation}
and bond current 
\begin{equation}
 |j^{\mathrm{PI}}|=2 J n^{\mathrm{PI}} \sin \Phi^{\mathrm{PI}}.
 \label{eq:picurrent}
\end{equation}
The maximal current $j^{\mathrm{PI}}= 8 J F^2/\gamma_b^2 $ occurs at $\Delta=-2 J\left(1+\cos \Phi^{\mathrm{PI}}\right)$, and the highest ratio $j^{\mathrm{PI}}/(J n^{\mathrm{PI}})=2 $ is found at $\Phi^{\mathrm{PI}} = \pi/2 $.

\begin{figure}[!t]
\includegraphics[width=\linewidth]{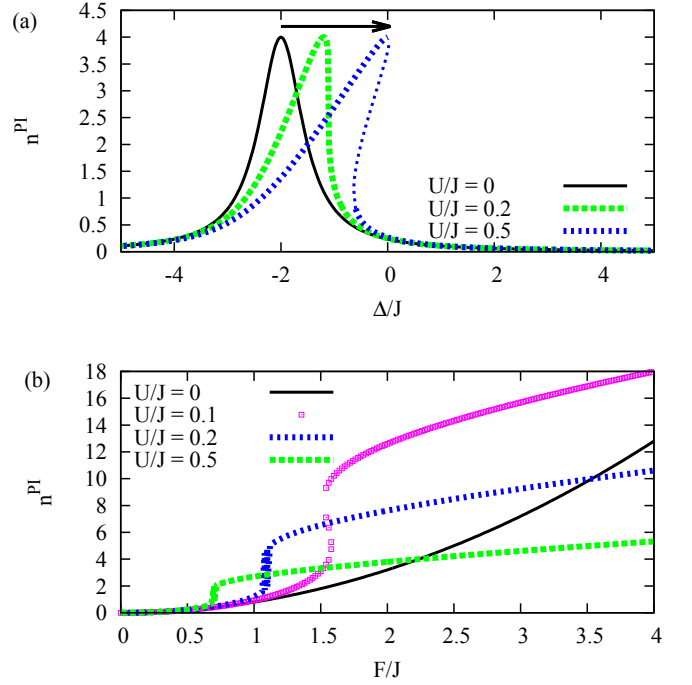}
\caption{(Color online) The density $n^{\mathrm{PI}} $ (\ref{eq:npiweaku}) as a function of a) detuning and b) driving for the setup (\ref{eq:phaseimprint}). Additional parameters used in the calculations: $\Phi^{\mathrm{PI}} = \pi/2 $, (a) $F/J=1$,  $\gamma_b/J = 1$, (b) $\Delta/J = -1$, $\gamma_b/J = 1$. The thin part of the dotted line in (a) corresponds to unstable solutions.}
\label{Fig:FigMH}
\end{figure}

We now discuss effects of weak interactions on the currents for $\Phi^{\mathrm{PI}}  = \frac{\pi}{2} $. The lattice density is obtained from Eq.~(\ref{eq:GP}) by solving
\begin{equation} 
 n^{\mathrm{PI}} = \frac{F^2}{\left( -2 J -\Delta+n^{\mathrm{PI}} U\right)^2+\frac{\gamma_b^2}{4}}, 
 \label{eq:npiweaku}
\end{equation}
while the bond current is still given by Eq.~(\ref{eq:picurrent}). From Eq.~(\ref{eq:npiweaku}) it is clear that the maximal current is the same as without interactions, only the resonance condition is changed to
\begin{equation}
\Delta^{\mathrm{PI}}_{\mathrm{r}} = -2 J + 4 U \frac{F^2}{\gamma_b^2}. 
\label{eq:resMH}
\end{equation}
This effect is illustrated in Fig.~\ref{Fig:FigMH}(a), where we also see that in certain regimes the mean--field description predicts up to three solutions for the same detuning $\Delta$ \cite{Drummond}. 
From Fig.~\ref{Fig:FigMH}(b) it is evident that only in the limit of low filling we find $j\sim F^2$ as in the case of $U=0$. At a certain threshold value of $F$, the dependence becomes steep and finally turns into $j \sim F^{2/3}$. We note that even stronger switching from low to high occupation can be found for nonlinear waveguide where normal modes synchronize during this switching process \cite{Leib2014}.

\begin{figure}[!h]
\includegraphics[width=\linewidth]{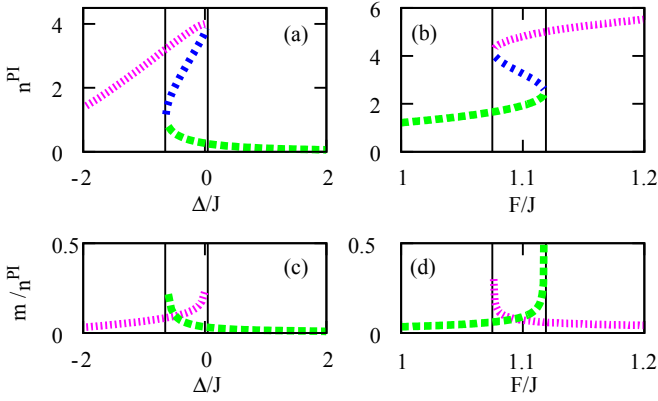}
\caption{(Color online) Top: The density $n^{\mathrm{PI}} $  as a function of (a) detuning and (b) driving for phase imprinting $\Phi^{\mathrm{PI}} = \pi/2 $. Bottom: quantum fluctuations $ m/n^{\mathrm{PI}}$. Additional parameters used in the calculations: (a), (c) $F/J=1$,  $\gamma_b/J = 1$, $U/J = 0.5$, (b), (d) $\Delta/J = -1$, $\gamma_b/J = 1$, $U/J = 0.2$. }
\label{Fig:FigMH2}
\end{figure}

By inspecting the contribution of quantum fluctuations given in Eq.~(\ref{eq:qfqf}) for different solutions (\ref{eq:npiweaku}), we find that in the region of coexistence one branch of solutions is unstable \cite{Drummond, Carusotto3} (the blue (middle) curve in Figs.~\ref{Fig:FigMH2}(a) and (b)). The two other branches exhibit stronger fluctuations in the intermediate regime, see  Figs.~\ref{Fig:FigMH2}(c) and (d), indicating that the accuracy of the mean-field approach deteriorates and the exact solution may be a superposition of the two  mean--field solutions. This conclusion is in agreement with  a variational solution of Eq.~(\ref{eq:eom}) that captures beyond mean--field effects and exhibits a unique steady state \cite{Weimer}.

In the limit of stronger interactions, in the Gutzwiller mean--field description (\ref{eq:gw}) our system decomposes into single cavities with an effective driving
\begin{equation}
 \eta = F -2 J \phi^{\mathrm{PI}}\left(1+ \cos \Phi^{\mathrm{PI}}\right),
\end{equation}
which incorporates contributions from the nearest--neighbors of every site of the square lattice. Our numerical results 
can be explained using an analytical result of Drummond and Walls \cite{Drummond} for a steady state of a single driven cavity.
In the steady state regime the value of the bosonic coherence  $\phi^{\mathrm{PI}}$ satisfies the equation \cite{Drummond, Boite, Boite2}
\begin{equation}
\phi^{\mathrm{PI}}  = \frac{\eta}{\Delta+i\gamma_b/2}\times\frac{\mathcal{F}(1+c,c^{*},8|\eta/U|^2)}{\mathcal{F}(c,c^{*},8|\eta/U|^2)}.
\label{eq:hg}
\end{equation}
The average density is given by
\begin{equation}
n^{\mathrm{PI}} = \left|\frac{2\eta}{U}\right|^{2}\times\frac{1}{|c|^2} \times\frac{\mathcal{F}(1+c,1+c^{*},8|\eta/U|^2)}{\mathcal{F}(c,c^{*},8|\eta/U|^2)},
\end{equation}
where $c = -2(\Delta+i \gamma_b/2)/U$,
\begin{equation*}
\mathcal F(c, d, z) = \sum_n^{\infty} \frac{\Gamma(c)\Gamma(d)}{\Gamma(c+n) \Gamma(d+n)}\times\frac{z^n}{n!}
\end{equation*}
is the generalized hyper--geometric function and $\Gamma(x)$ is the gamma function.
Our analysis is analogous to the analysis performed in Refs.~\cite{Boite, Boite2}, with the main difference that we introduce the parameter $\Phi^{\mathrm{PI}}$, which is a necessary ingredient to obtain currents. The steady states we obtain by solving Eq.~(\ref{eq:hg}) are also found in real time evolution of Eqs.~(\ref{eq:rtgw}) starting from an initial state with a strong bosonic coherence.

\begin{figure}[!tbh]
\includegraphics[width=\linewidth]{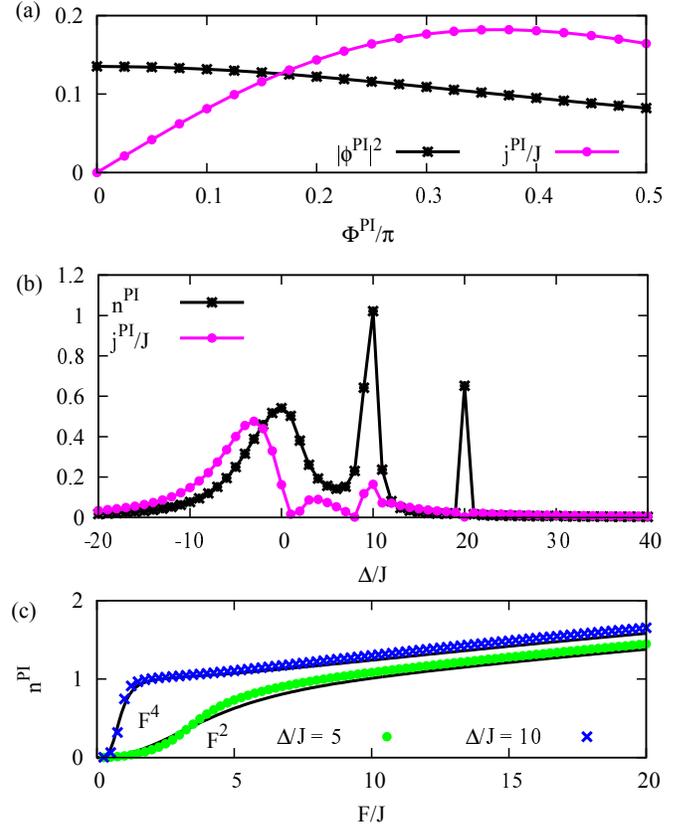}
\caption{(Color online) The  current $j^{\mathrm{PI}} = 2 J |\phi^{\mathrm{PI}}|^2 \sin \Phi^{\mathrm{PI}} $ and local density $n^{\mathrm{PI}}$  in the steady state of the phase imprinting setup (\ref{eq:phaseimprint}). Parameters: $U/J = 20$, $\gamma_b/J =0.2 $, (a) $\Delta/J = 10, F/J = 2.4, n^{\mathrm{PI}}\approx1$, (b) $F/J=2.4, \Phi^{\mathrm{PI}} = \pi/2$,  (c) $\Phi^{\mathrm{PI}} = \pi/2 $. The black solid lines in (c) are the corresponding analytical results for $J=0$. }
\label{Fig:FigHG}
\end{figure}
Our main results are summarized in Fig.~\ref{Fig:FigHG}. As the strong interaction $U/J=20$ tends to suppress bosonic coherences, the ratio of $j^{\mathrm{PI}}/(J n^{\mathrm{PI}}) $ is an order of magnitude smaller compared to the non--interacting regime. The maximal ratio is found at $\Phi^{\mathrm{PI}} \approx 0.35 \pi$, since the bosonic coherence $\phi^{\mathrm{PI}}  $ is higher for this value than at $\Phi^{\mathrm{PI}}  = \pi/2 $, Fig.~\ref{Fig:FigHG}(a).  The  current $j^{\mathrm{PI}} $ is a non--monotonous function of the detuning $\Delta$, Fig.~\ref{Fig:FigHG}(b). This behavior stems from multiphotonic resonances of the single cavity that occur at \cite{Drummond, Boite2}
\begin{equation}
 \Delta_{\mathrm{r}}^{\mathrm{PI}} = \frac{U}{2} (n-1), n=1,2,\ldots,
 \label{eq:dc}
\end{equation}
 when the energy of $n$ incoming photons is equal to the energy of $n$ cavity photons.
   \begin{figure}[!tbh]
 \includegraphics[width=\linewidth]{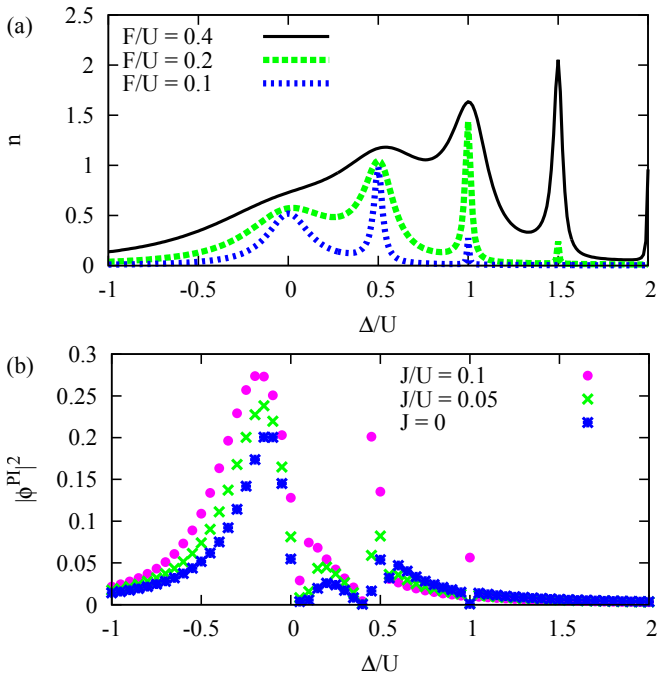}
\caption{(Color online) (a) Analytical results from Ref.~\cite{Drummond} for a steady state of a single driven cavity.  (b) Bosonic coherence as a function of the detuning for several values of $J$. Parameters $\gamma_b/U=0.01 $ and $F/U = 0.12, \Phi^{\mathrm{PI}} = \pi/2 $ in (b). }
\label{Fig:FigHGJ0}
\end{figure}
 The number of resonances that can be resolved practically is set by the ratio $F/U$, which also determines the maximal possible filling of the lattice. For very weak driving only low--lying resonances can be probed, as shown in Fig.~\ref{Fig:FigHG}(b) for $F/U=0.12 $. At  stronger driving, low--order resonances are washed out --- as can be seen from the analytical solution available for $J = 0$, see Fig.~\ref{Fig:FigHGJ0}(a) --- and replaced by a simpler dependence that is captured by Eq.~(\ref{eq:GP}). Yet, a few high--lying resonances can be resolved clearly even at strong $F$, see Fig.~\ref{Fig:FigHGJ0}(a). In the vicinity of the lowest--order resonance, maximal $j^{\mathrm{PI}} $ is found at some off--resonant negative value of $\Delta$, while higher--order resonances can appear either as peaks or dips in the current intensity. In Fig.~\ref{Fig:FigHGJ0}(b) we observe a local maximum of the coherence at  $\Delta = U/2$, while at $\Delta = U$ there is a minimum at $J/U = 0.05$ and maximum at $J/U = 0.1$. When $J/U$ and $F/U$ are comparable, a regime with multiple stable mean-field solutions can be found \cite{Boite, Boite2}, however, this topic is beyond the scope of this paper.

In order to infer the dependence of the current on the driving amplitude in the regime of strong $U$, we expand the analytical result \cite{Drummond} for $J=0$ in the limit of weak $F$ and obtain
\begin{eqnarray}
 n&\sim& \frac{1}{U^6} \left[\left(\gamma_b^2+4 \Delta^2\right) \left(\left(U-2 \Delta\right)^2+\gamma_b^2\right) F^2\right.\nonumber\\
 &+&\left.8 U \left(4 \Delta-U\right) F^4+\ldots\right].
\end{eqnarray}
If the dissipation rates are low ($\gamma_b/U\ll 1$), at $\Delta = U/2$ the term proportional to $F^4$ will dominate the $F^2$ term even at very weak $F$, as we clearly observe in Fig.~\ref{Fig:FigHG}(c) at $\Delta/J = 10, U/J = 20.$ Except for this special resonant case, we typically have an $F^2$ dependence in the weak $F$ limit. In the regime of strong $F$, we recover the result obtained in the previous section $j \sim F^{2/3}$.

\subsection{Source--drain setup}

\begin{figure}[h]
    \includegraphics[width=\linewidth]{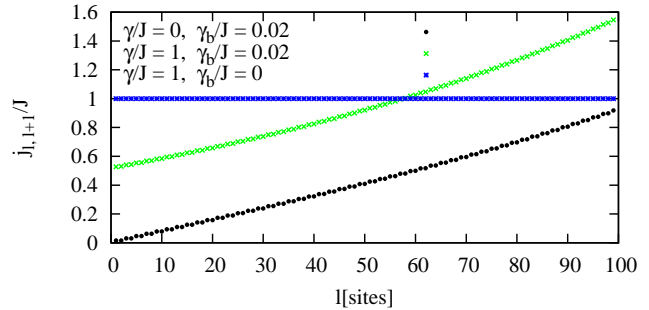}
    \caption{(Color online) The currents $j_{l,l+1}$ between nearest--neighbor sites along $x$-direction for the source--drain setup (\ref{eq:sourcedrain}). Parameters used: $F/J = 1 $, $\Delta/ J = -2$.}
\label{Fig:Figcdist}
\end{figure}

 Typical spatial distributions of the bond currents in the non--interacting regime of the setup defined in Eq.~(\ref{eq:sourcedrain}) are presented in Fig.~\ref{Fig:Figcdist} for a lattice size of $N=100$ in $x$-direction and assuming translational invariance in $y$-direction, where a single site is repeated periodically. The driving is applied at the rightmost lattice sites and in the presence of uniform dissipation rates, the intensity of the bond currents decays roughly linearly as we approach the leftmost sites. In order to enhance overall currents, we consider the leftmost cavities to exhibit a stronger dissipation rate. In the idealized case of $\gamma_b = 0$ we find a uniform current throughout the lattice. Hence, in the following we will explore the source--drain (SD) setup Eq.~(\ref{eq:sourcedrain})
with open boundary conditions in $x$- and periodic boundary conditions in $y$-direction. The differences of this setup with respect to the model studied in Ref.~\cite{Biella} are the following: we consider a two--dimensional lattice and we take into account spatially varying dissipation rates of cavities, see Eq.~(\ref{eq:sourcedrain}). Moreover, we investigate a regime of high lattice density and weak interactions, which was not addressed in Ref.~\cite{Biella}.

In the steady state regime with constant total number of photons, it holds true that
\begin{equation}
 -2 F\, \mathrm{Im}\phi_N = \gamma n_1 + \gamma_b \sum_{j=1}^N n_{j},
 \label{eq:equilibriuminh}
\end{equation}
i.e.~the flux of incoming particles on the right is equal to the flux of the particles leaving the system (continuity equation).
In the special case of $\gamma_b = 0$ we find a uniform current
\begin{equation}
j^{\mathrm{SD}} = \gamma n_1   =-2 F \mathrm{Im}\phi_N.
 \label{eq:equilibrium}
\end{equation}

 \begin{figure}[!tbh]
 \includegraphics[width=\linewidth]{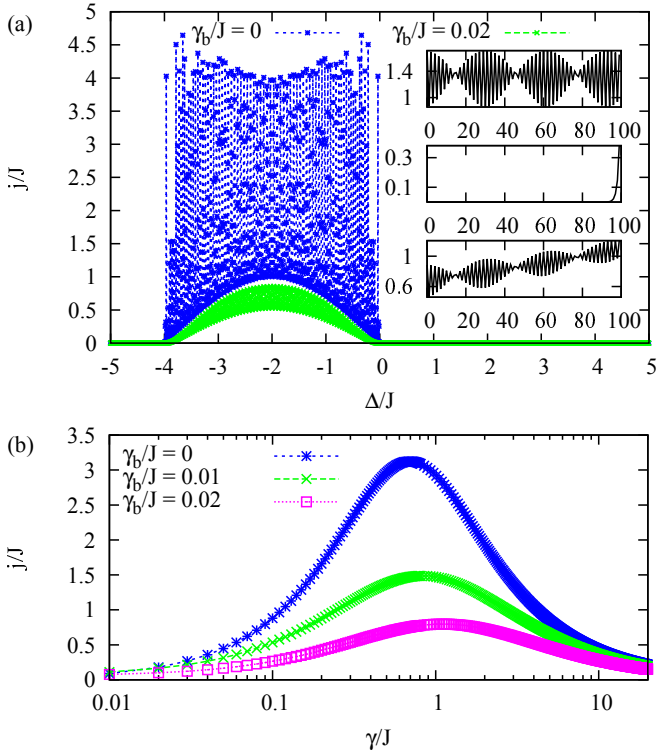}
\caption{(Color online) The current (\ref{eq:current}) for the setup (\ref{eq:sourcedrain}) as a function of (a) detuning and (b) local dissipation. Additional parameters: a) $\gamma/J=1$, $F/J=1$, b) $\Delta/J = -2.1$, $F/J=1$. 
Insets in a) show the spatial distribution of $|\phi_l|$ over the site index $l$ in $x$-direction. Typical distributions range from (top) ``conducting'' behavior ($\Delta/J = -2.1, \gamma_b = 0$) to (center) the situation without bulk current ($\Delta/J = 5, \gamma_b = 0.0$). (bottom) ``Conducting'' behavior with bulk dissipation ($\Delta/J = -2.1, \gamma_b/J = 0.02$). The lattice consists of $N=100$ sites in $x$-direction.}
\label{Fig:Fig1}
\end{figure}
In the non--interacting limit of the setup (\ref{eq:sourcedrain}), both the total density $\sum_l \langle n_l \rangle$ and the intensity of the bond current are proportional to $F^2$ according to Eq.~(\ref{eq:inverse}).  In  Fig.~\ref{Fig:Fig1}(a) we show that the transport occurs if there is an eigenmode of $H^0$ in Eq.~(\ref{eq:h0}) at the given value of $\Delta$ to support it.  In our case the range of resonant driving frequencies is  $\Delta\in[-4J, 0]$, as the frequency of the lowest mode of a two dimensional lattice is $-4 J$ and we only consider transport in $x$ direction. To infer effects of local dissipation $\gamma$, we invert the matrix $M$ (\ref{eq:matrixm}), first for $\gamma_b = 0$. The bond current is given by
\begin{equation}
j = \gamma \frac{F^2}{J^2} \frac{p\left(2+\frac{\Delta}{J}\right)}{q\left(2+\frac{\Delta}{J}\right)+\frac{\gamma^2}{4 J^2}r \left(2+\frac{\Delta}{J}\right)}, 
\end{equation}
where $p(x)$, $q(x)$, $r(x)$ are polynomials that can be expressed in terms of determinants of the matrix $M$ and its sub-matrices with $\gamma$ set to zero.
The last dependence is plotted in  Fig.~\ref{Fig:Fig1}(b) and we see that the bond current is maximal when the dissipation rate $\gamma$ is of the same order of magnitude as the hopping rate $J$, i.e.~$\gamma/J\sim 1$. Beyond this value, the current is suppressed as the quantum Zeno effect takes place \cite{Kordas}. If the resonant condition $\Delta = \varepsilon_n $ is fulfilled, the matrix M in Eq.~(\ref{eq:matrixm}) is singular for vanishing $\gamma_b$ and we find $j\sim\gamma^{-1}$. 
As expected, the intensity of $j$ is suppressed by the presence of finite bulk dissipation $\gamma_b$. In Fig.~\ref{Fig:Fig1}(b) at finite $\gamma_b$ we plot the current between the leftmost site and its nearest neighbor in $x$-direction. The insets of Fig.~\ref{Fig:Fig1}(a) show density distributions in different regimes. In the conducting regime density profiles are typically non--uniform.

\begin{figure}[!t]
\includegraphics[width=\linewidth]{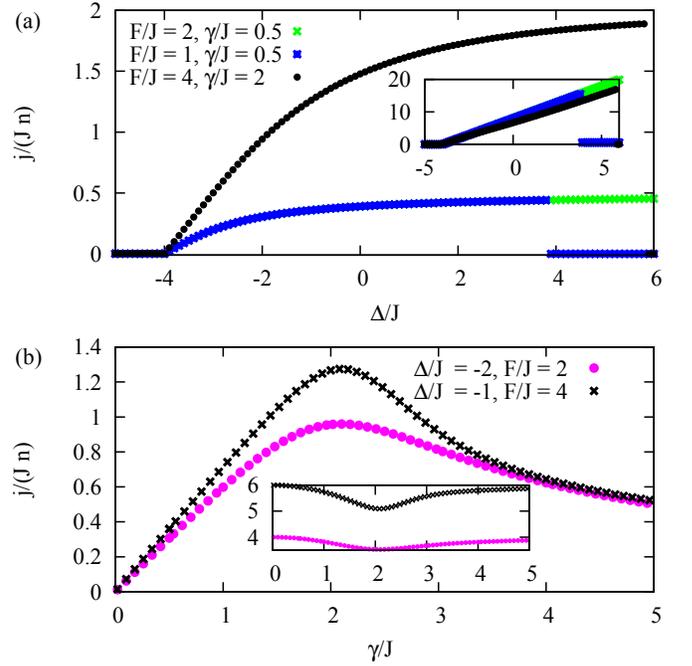}
\caption{(Color online)  The ratio $j/(J n)$ as a function of (a) detuning $\Delta$ and (b) the dissipation $\gamma$ for $U/J=0.5$, $\gamma_b=0$, and $N = 50$ for the source--drain setup (\ref{eq:sourcedrain}). Insets of both plots show the local density $n$ in the bulk. }
\label{Fig:Figwicon}
\end{figure}
Now we address effects of weak interactions first with $\gamma_b = 0$. To access the steady states, we perform a real--time propagation of Eq.~(\ref{eq:GP}). This method raises an important question about if and how the steady states depend on the chosen initial conditions \cite{Spohn}. For very weak $U$, such that $nU/J\ll1$, the non--interacting steady states from the previous section provide a good starting point. States obtained in this way exhibit non--uniform density distributions. As $U$ becomes stronger, our numerical results suggest that in the bulk of the system, where $\gamma_l = 0$ and $F_l = 0$, the steady states are given by $\phi_l = \sqrt{n^{\mathrm{SD}}}\exp(i \Phi l) $. The density is uniform in the bulk
\begin{equation}
n^{\mathrm{SD}}(\Phi) =\frac{\Delta+2 J(1+\cos \Phi)}{U},
\label{eq:unn}
\end{equation}
and so is the bond current
\begin{equation}
 j^{\mathrm{SD}}(\Phi) = 2 Jn^{\mathrm{SD}}(\Phi) \sin \Phi,
 \label{eq:unj}
\end{equation}
where $\Phi$ is a constant phase difference between $\phi_l$ of nearest neighbors. Unlike the phase imprinting setup, where the value of $\Phi$ is fixed by the external drive, here the phase difference is set by the boundary conditions (\ref{eq:equilibrium}). In the following, we set the initial state for the real time propagation of Eq.~(\ref{eq:GP}) to a steady state for fixed values of $\Delta$, $F$ and $\gamma$, then adiabatically change one of the parameters and monitor how this change affects the steady state.

\begin{figure*}[!t]
	\includegraphics[width=\linewidth]{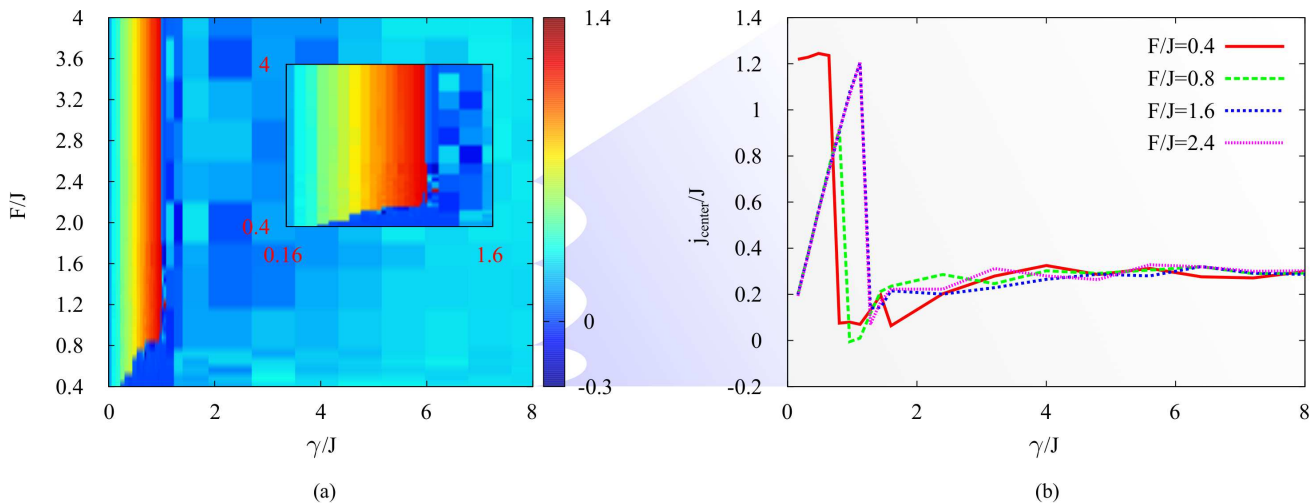}
	\caption{(a) Average current $j/J$ in the center of a system \eqref{eq:sourcedrain} of width $N=200$ sites in $x$-direction and translational invariance in $y$-direction at times $tJ\sim 10^4$. At $\gamma_c/J \sim 1$ the current is suppressed due to the quantum Zeno effect. In (b) slices through the phase diagram at different driving strengths are shown. Parameters used are $U/J=4,\Delta/J = 2$.}
	\label{fig:zeno_phase_diagram}
\end{figure*}
As in the phase imprinting setup, for very weak $U$ it holds that $j\sim F^2$. On the contrary, in the steady state (\ref{eq:unn}) the driving $F$ affects only the rightmost sites and not the bulk features. As $F$ gets smaller, only the occupancy of the rightmost sites $n_{N}$  decreases. Eventually, densities on the leftmost and rightmost lattice site become equal $n_{N}\approx n_{1}$ and at this point the steady state is no longer supported. 
This occurs approximately at $F^*=\frac{1}{2}\gamma \sqrt{n_{1}}$ and we have $j \sim \theta(F-F^*) $, where $ \theta(x)$ is a step function. With further decrease of the driving intensity $F$, our numerical results exhibit strong oscillations that persist up to the longest integration time. In this regime, numerical simulations fail to converge to a stationary regime and the average intensity of the bond current is zero.

The steady states (\ref{eq:unn}) exist if $\Delta \geq -4 J$. Above this threshold the lattice filling exhibits a roughly linear increase with $\Delta$. The detuning also affects the phase difference $\Phi$, as evidenced by the change in the ratio $j/n$, see Fig.~\ref{Fig:Figwicon}(a). The current per particle saturates at large $\Delta$ and it turns out that at large enough $\Delta$, when the lattice filling is too high, the steady state is no longer supported for it requires stronger driving $F$.

In the source--drain setup the value of $\Phi$ can be changed by tuning the intensity of the local dissipation $\gamma$ \cite{Zezyulin}.
Unlike $F$, $\gamma$ affects both the bulk density of a steady state  as well as the strength of the bond current. For example, in the case presented in Fig.~\ref{Fig:Figwicon}(b) an optimal ratio $j/(J n) \approx 1$ is found at  $\gamma/J\approx  2$. By additionally optimizing the detuning $\Delta$, this ratio can be enhanced further, see Fig.~\ref{Fig:Figwicon}(a). In a similar way as for the phase imprinting, effects of quantum fluctuations can be estimated and we find them to be reasonably small. Finally, we find that the states (\ref{eq:unn}) are stable with respect to the bulk dissipation for moderate values of $\gamma_b/J\sim0.01$. 


We now investigate features of the current for stronger interactions at a fixed ratio $U/J = 4$ as a function of the external system parameters $\gamma, F$ ($\gamma_b = 0$) by solving Eq.~\eqref{eq:rtgw} for long times.
In Fig.~\ref{fig:zeno_phase_diagram} we show the average current at large times $tJ\sim 10^4$, where we identify quasi--steady states, which yield approximately constant current and particle densities $j,n \approx \mathrm{const}$.  We average theses quantities over a large enough time span, which evens out most of the oscillations, and we attribute any residual noise to lower--frequency components, which stem from our choice of the initial state. At small $\gamma/J \lesssim 1$ the aforementioned quasi--steady states exist and their current density increases almost linearly with $\gamma/J$. The current density is then only weakly dependent on the driving $F/J$. At $\gamma_c/J \sim 1$ a sharp transition occurs and the existence of the quasi--steady states is suddenly violated. What we find instead are oscillating mixed states with (almost) vanishing average current density, hence a non--transparent region. 

\begin{figure*}[!t]
	 \includegraphics[width=\linewidth]{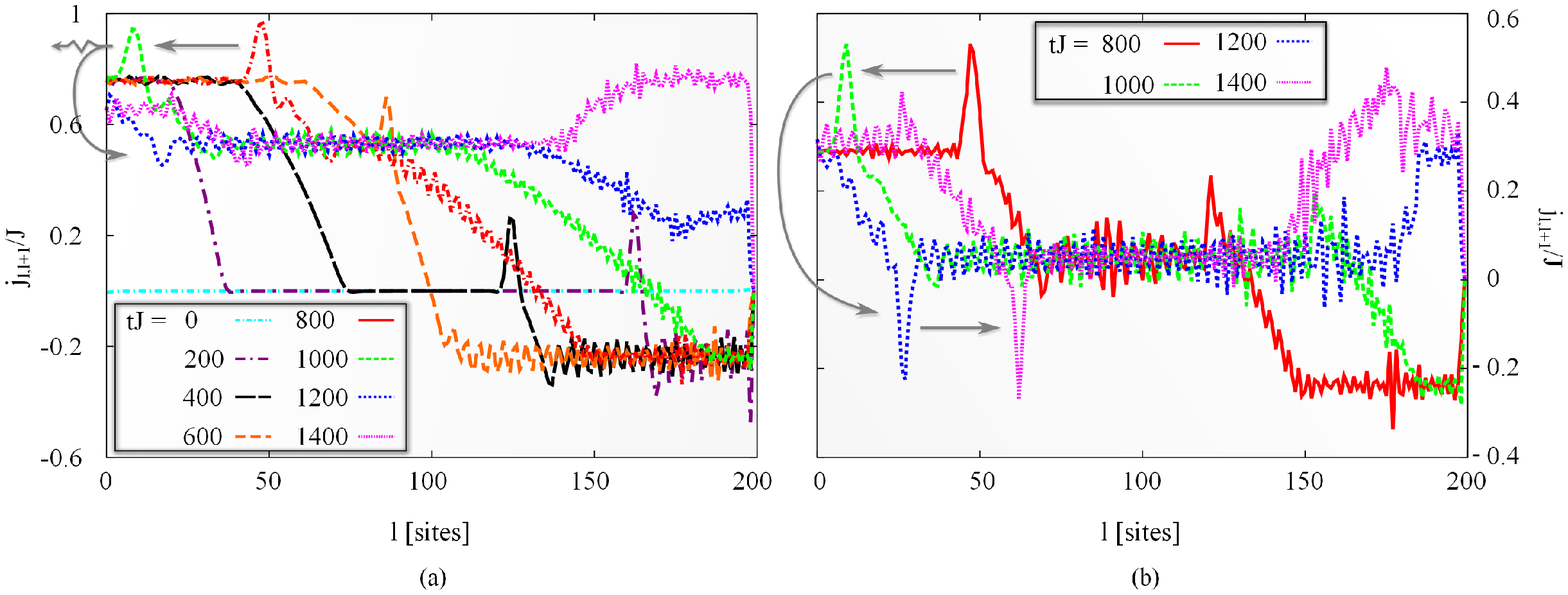}
	\caption{Current distribution $j_{l,l+1}/J$ at small times $tJ\in \lbrack 0,1400\rbrack$ for the system of \figref{fig:zeno_phase_diagram} at (a) $\gamma/J = 0.8$ and (b) $\gamma/J = 4$. In (b) currents are reflected from the dissipative site as a consequence of quantum Zeno blocking. The initial dynamics for $tJ\in \left[0,600\right]$ are only shown in (a) for clarity. Relevant are the peaks in the current distribution, arrows are meant to guide the eye. Parameters used are $U/J=4,\Delta/J=1, F/J = 2$.}
	\label{fig:current_reflection}
\end{figure*}
We explain this observation with the quantum Zeno effect \cite{CaldeiraLeggett, CiracSchenzle} by identifying the loss rate $\gamma$ with the rate of a generalized measurement, which --- repeated at high frequencies --- stops the unitary time evolution and forces the system into the lossless steady state, where no significant particle transfer from the driven to the lossy site is observed. Following early theoretical considerations \cite{Misra}, the quantum Zeno effect was observed in experiments with cold ions \cite{Itano} and ultracold atoms \cite{Rempe_quantum_zeno, Nagerl_quantum_zeno, BarontiniLabouvie, Zhu}. In the context of ultracold atoms, the interplay of interactions and dissipation has received a lot of attention \cite{Vardi, sh1, sh2, TrimbornWitthaut, Kollath, Kollath2, Vidanovic, Rajagopal}.
Applying this principle to our system we first note that only the dissipative sites (in this case only the ones at the left boundary $x=0$) are being ``measured'', which means that only the reduced density operators on these sites become time--independent in the limit of frequent measurements, i.e.~strong dissipation. In fact, in the limiting case the local density operators will be equal to the local vacuum. The rest of the system will henceforth pursue it's own unitary time evolution, where the coupling to the dissipative site is simply disabled. This explains why we cannot find quasi--steady states at large dissipation, because the only steady states under unitary time evolution are eigenstates of $\mathcal{H}$ and for arbitrary initial states, composed of many different eigenstates of $\mathcal{H}$, observables do not converge.

The transition at $\gamma_c/J \sim 1$ occurs at the point where the time scales of the local measurement $\sim 1/\gamma$ and the competing hopping process at rate $J$ are balanced. At this point the current/particle transfer is maximal since particle loss occurs at the same rate as the hopping, which fills up the dissipative sites again. If the dissipation is any stronger this filling process will be suppressed.

From this discussion it is already apparent that the dissipation is the prevalent ingredient for a description of the transport in this system. 
Microscopically, this can be understood from a wave picture, where excess currents are reflected from a hard wall and destructive interference of counter--propagating waves takes place. We confirm this assumption by examining snapshots of the current distribution at small times, see Fig.~\ref{fig:current_reflection}, before the quasi--steady state regime has been reached. By observing the time evolution of easily identifiable current peaks we find that for weak $\gamma$ only a small proportion of particles is reflected while the majority is transmitted to the lossy site and lost eventually, Fig.~\ref{fig:current_reflection}(a). However, for a large enough ratio $\gamma/J$, currents are reflected --- not at the system boundary, but at the lossy site, Fig.~\ref{fig:current_reflection}(b). Peaks traveling towards the dissipative edge will change the direction, i.e.~the sign of the current, just before the dissipative site.  As a consequence the dissipative site is effectively decoupled from the system.

The oscillations in the region with $\gamma_c < \gamma \ll \infty$ can  be explained in the wave picture as well. Since perfect destructive interference of reflected components would require suitable geometric conditions, which we do not alter throughout our simulations, the process of particles ``bouncing'' back and forth will lead to a small current distribution, which is difficult to average out completely.

The source--drain setup \eqref{eq:sourcedrain} is the simplest way to describe transport through the system, neglecting the penetration depth of the laser into the medium and de-excitations in the bulk. Typically, lattices are formed of identical cavities, whose individual mode excitations have the same decay rates, so that a constant bulk decay rate is more realistic. In order to simulate the penetration of the laser into the medium we consider a decaying laser amplitude $F$ as a function of the penetration depth. In the simplest case this would be a linear decay with bulk dissipation present:
\begin{equation}
F_l = F_0 +\Delta F \left(l-1\right) , \quad \gamma_l = \gamma,
\label{eq:intensitygradient}
\end{equation}
 where $l$ denotes the site index in $x$-direction and no explicit $y$-dependence is given, as before.

\begin{figure*}
	\includegraphics[width=\linewidth]{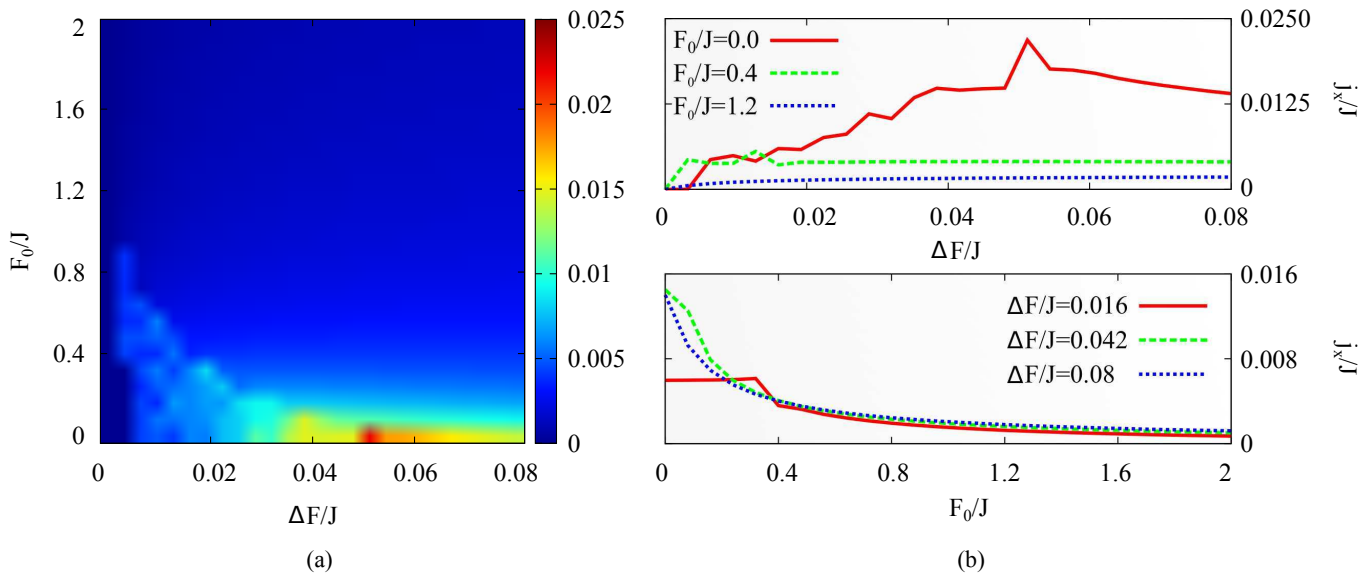}
	\caption{(a) Average current $j_x = 1/N \sum_l j_{l, l+1}$ as a function of the slope $\Delta F$ of the driving laser amplitude \eqref{eq:intensitygradient} and the background amplitude $F_0$. Here, the system has a width of $N =200$ sites in $x$-direction and one site in $y$-direction, which is repeated via periodic boundary conditions. Cuts through the diagram are shown in (b), where we observe maximal current at a specific value of the slope $\Delta F$ in the upper plot for small $F_0$. The maximum is shifted to the left with increasing $F_0$. Varying $F_0$ at fixed $\Delta F$ (lower plot) shows decreasing behavior of $j_x$. Parameters used are $U/J = 4, \Delta/J =1, \gamma/J = 0.2$.}
	\label{fig:amplitude_driveU}
\end{figure*}

For the setup (\ref{eq:intensitygradient}) we investigate  the dependence of the currents on an overall laser field $F_0$  and an ``on top" gradient $ \Delta F$. It turns out that the larger the offset field  $F_0$, the lower the overall current, Fig.~\ref{fig:amplitude_driveU}(a). In \figref{fig:amplitude_driveU}(b) we observe a peak in the photon transport at $F_0=0$ and a strongly suppressed transport for any other value of $F_0$.  The effect can be understood by realizing that the off-set field corresponds to the phase imprinting with phase zero, i.e. we are pumping a mode that doesn't support any current.
Effects of the gradient $\Delta F$ are given in \figref{fig:amplitude_driveU}(b). At low $\Delta F$, the system compensates for the imbalance between neighboring sites via coherent transport of photons along the gradient and as expected, the gradient enhances the current. However, at a certain value of the $\Delta F$ the imbalance is so strong that the incoherent dynamics becomes the dominant process.

\section{Conclusions}
\label{Sec:Conclusions}

Motivated by ongoing research interest in arrays of coupled photonic cavities, we have investigated different possibilities to optimize coherent transport in this setup. We have started from the non--interacting limit, where simple relations between the bond current and externally tunable parameters can be established. To address the role of interactions we have employed the Gutzwiller mean--field theory and a simpler Gross--Pitaevskii--like approach when possible.

In the case where bond currents are introduced by phase engineering of the external lasers, we have found that weak interactions shift the driving frequency that leads to a peak in the current toward higher values. On the other hand, in the strongly interacting regime of this setup, multiphotonic resonances of a single driven cavity lead to multiple peaks of the current as a function of the driving frequency. The lattice filling is set by the strength of the applied driving field $F$ and the dissipation rate $\gamma_b$, but interactions can modify the $F^2$ proportionality into either a weaker $F^{2/3}$ gain or into an effectively stronger gain in the vicinity of multiphotonic resonances. 

In the source--drain setup, local dissipation $\gamma$ proves to be the tuning parameter that allows to maximize the bond current. The optimal value of $\gamma$ is set by the intrinsic hopping rate of the underlying Bose--Hubbard model. Further increase of $\gamma$ leads to the quantum Zeno dynamics that suppresses uniform currents. The effects of the applied driving $F$ turn out to be especially simple in the interacting case: the steady state is either stable at the specific value of $F$ or its stationarity breaks down as stronger driving strength would be required to balance the dissipation.

The main approximation of our analysis is the employed mean-field approach together with the simplified form of the bond current, that limits to the regime of the strong bosonic coherences. The contribution of non--trivial correlations becomes important in the limit of very strong interactions and weak driving and this case should be treated in the future using beyond mean-field approximations \cite{Schollwock, Daley, Hartmann, Biella, Weimer}. However, we expect that the main effects we have identified at finite coherences will not be modified by the inclusion of higher order terms. Another interesting research direction would be to connect our results with well--established results describing heat transport on the microscopic level \cite{Benentireview}.

\section{Acknowledgments}
The authors thank Daniel Cocks and Giuliano Orso for useful discussions.
Support by the German Science Foundation DFG via Sonderforschungsbereich SFB/TR 49, Forschergruppe FOR 801 and the high-performance computing center LOEWE-CSC is gratefully acknowledged.  
This work was supported in part by DAAD (German Academic and Exchange Service) under project BKMH.
I.~V.~acknowledges support by the Ministry of Education, Science, and Technological Development of the Republic of Serbia under projects ON171017 and BKMH, and by the European Commission under H2020 project VI-SEEM, Grant No. 675121. Numerical simulations were partly run on the PARADOX supercomputing facility at the Scientific Computing Laboratory of the Institute of Physics Belgrade.
M.~J.~H.~acknowledges support by the DFG via the grant HA9953/3-1  and by the UK Engineering and Physical Sciences Research Council (EPSRC) under EP/N009428/1.

\end{document}